\documentclass[preprint,superscriptaddress,prb,preprintnumbers,amsmath,amssymb]{revtex4}

\usepackage{graphicx,amssymb}

\begin{document}

\title{Genetic algorithms predict formation of exotic ordered\\ configurations
for two-component dipolar monolayers \\}

\author{Julia Fornleitner} 
\affiliation{Center for Computational Materials Science and Institut
f\"ur Theoretische Physik, Technische Universit\"at Wien, Wiedner
Hauptstra{\ss}e 8-10, A-1040 Vienna, Austria} 

\author{Federica Lo Verso}
\affiliation{Institut f\"ur Theoretische Physik II: Weiche Materie,
Heinrich-Heine-Universit\"at D\"usseldorf,
Universit{\"a}tsstra{\ss}e 1,
D-40225 D\"usseldorf, Germany}
\affiliation{D{\'e}partement de chimie min{\'e}rale, analytique et
appliqu{\'e}e,
Universit{\'e} de Gen{\`e}ve - Sciences II\\
30, Quai Ernest-Ansermet,
CH-1211 Gen{\`e}ve 4, Switzerland}

\author{Gerhard Kahl}
\affiliation{Center for Computational Materials Science and Institut
f\"ur Theoretische Physik, Technische Universit\"at Wien, Wiedner
Hauptstra{\ss}e 8-10, A-1040 Vienna, Austria}

\author{Christos N.~Likos}
\affiliation{Institut f\"ur Theoretische Physik II: Weiche Materie,
Heinrich-Heine-Universit\"at D\"usseldorf,
Universit{\"a}tsstra{\ss}e 1,
D-40225 D\"usseldorf, Germany}
\affiliation{The Erwin Schr{\"o}dinger International Institute for
Mathematical Physics (ESI), Boltzmanngasse 9, A-1090 Vienna, Austria}

\date{\today}

\begin{abstract}
We employ genetic algorithms (GA), which allow for an unbiased
search for the global minimum of energy landscapes, to identify
the ordered equilibrium configurations formed by binary dipolar systems
confined on a plane. A large variety of arrangements is identified,
the complexity of which grows with increasing asymmetry between the
two components and with growing concentration of the small particles.
The effects of the density are briefly discussed and a comparison
with results obtained via conventional lattice-sum minimization
is presented. Our results can be confirmed by experiments involving
Langmuir monolayers of polystyrene dipolar spheres or 
super\-para\-magnetic colloids confined on the air-water interface
and polarized by an external, perpendicular magnetic field.
\end{abstract}


\maketitle

\begin{figure*}[h]
      \begin{center}
      \begin{minipage}[t]{17.2cm}
      \includegraphics[angle=270,width=7.0cm ] {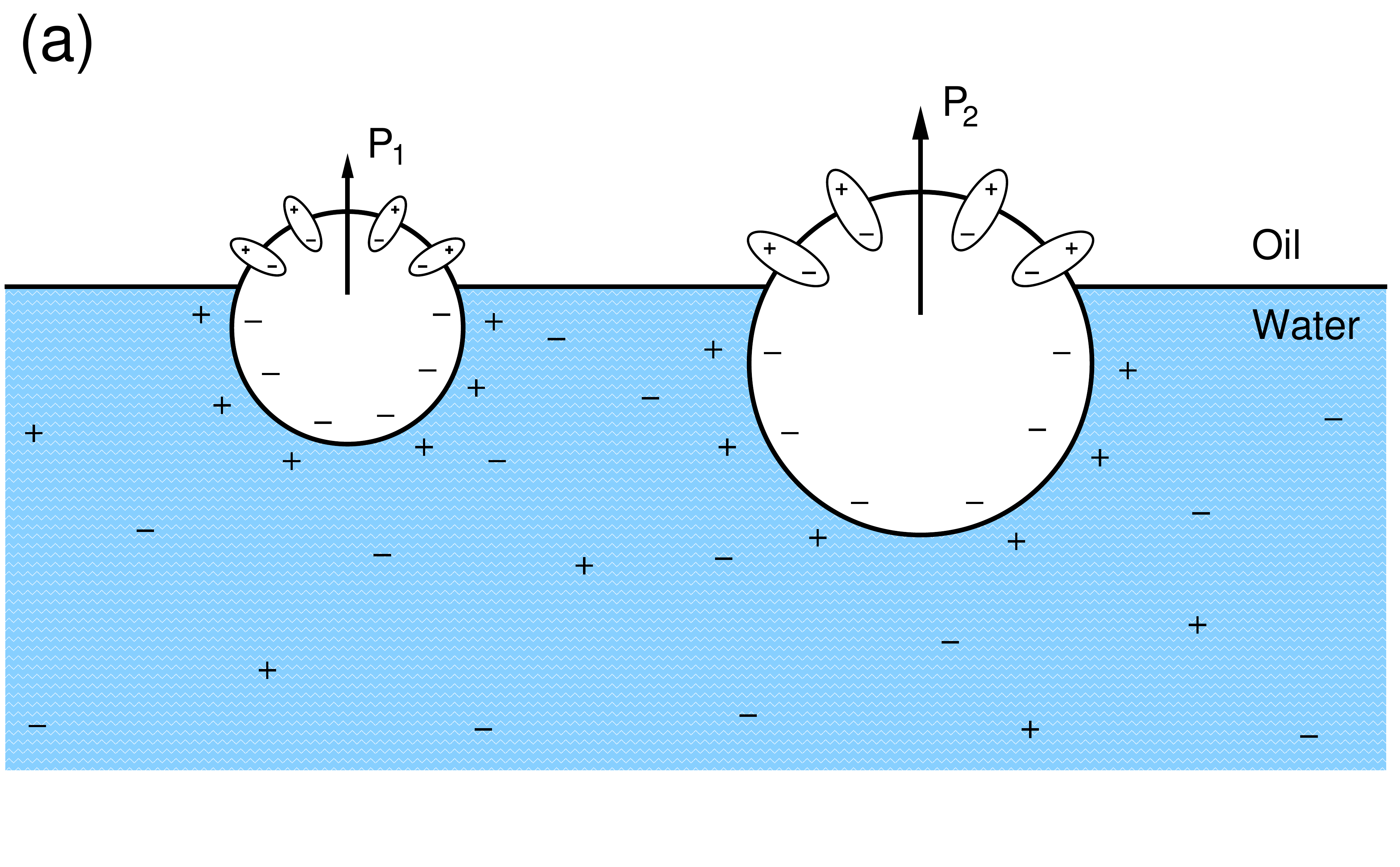}
      \hspace{0.5cm}
      \includegraphics[angle=270,width=7.0cm] {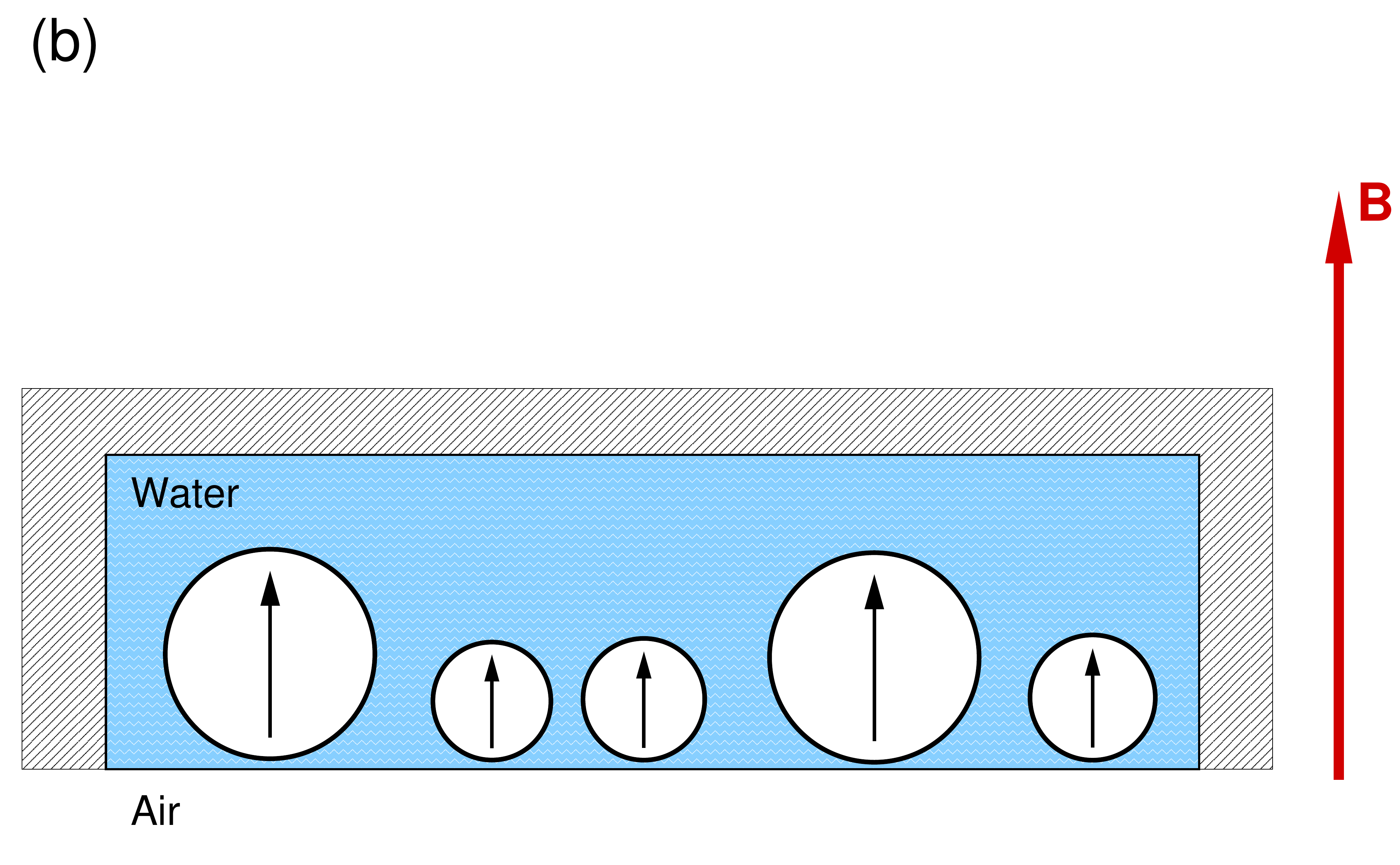}
      \end{minipage}
      \end{center}
\caption{(a) Schematic representation of system I (inspired by Ref. \onlinecite{Sti05}): 
  two particles of different size floating
  at an oil-water-interface. The dipole moments $P_1$ and $P_2$ are given by the
  vector sum of the dipole moments on the particle-oil-interface, since the dipole
  interaction is screened in the aqueous phase. (b) Schematic representation of system II:
  superparamagnetic colloidal particles trapped on the water/air interface in
  a pendant water droplet. The external magnetic field ${\bf B}$ is used to tune the interactions between the spheres.}
  \label{fig_schematic_monolayer}
\end{figure*}

Investigations of the structural and thermodynamic properties of
colloids confined at fluid interfaces represent a very active
topic of current research. Unlike particles in the bulk, the effective
interactions between the colloidal particles at fluid interfaces are influenced not
only by the properties of the particles and the solvent but, in addition,
by the surface and line tensions of the interface.\cite{bresme:jcpm:07}
Here we focus on the self-assembly scenarios of binary dipolar colloids, a system
for which various experimental realizations exist, and whose interactions
have been quantitatively established.
Experimentally, studies of two-dimensional ordered arrangements of
colloidal particles can be fairly easily realized by investigating polystyrene
particles floating on an oil-water interface\cite{Ave00LAN16a,Ave00LAN16b}
(system I). Their size, typically lying in the micrometer domain,
allows a direct observation of the particles in light microscopes. Computer
simulations for the binary case,\cite{Sti05} mimicking the experimental setup 
sketched in Fig.~\ref{fig_schematic_monolayer}(a), revealed a surprisingly rich
spectrum of exotic ordered equilibrium structures. Another realization 
of such systems is offered by the setup of Maret and 
coworkers,\cite{maret:prl:97,maret:prl:99,maret:prl:00,maret:prl:03,maret:prl:06}
employing
superparamagnetic colloids suspended on a pendant water droplet,
as sketched in Fig.~\ref{fig_schematic_monolayer}(b) (system II).
Here, a magnetic field ${\bf B}$ applied in the perpendicular direction polarizes the particles and induces a repulsive
interaction among them that scales with interparticle separation
$r$ as $\sim r^{-3}$, see Refs.\ \onlinecite{maret:prl:06,likos:jpcm:06}.   

\begin{figure*}[h]
      \begin{center}
      \begin{minipage}[t]{17.2cm}
      \includegraphics[width=7.0cm, clip] {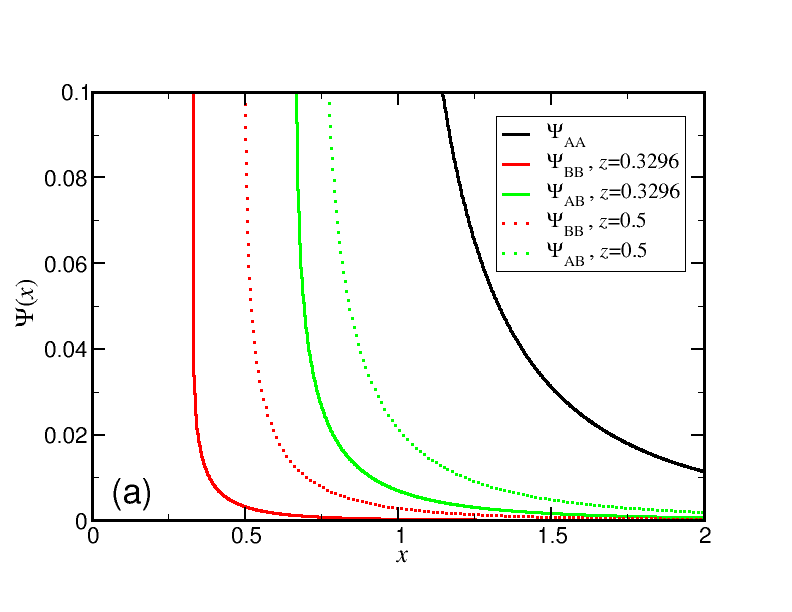}
      \hspace{0.5cm}
      \includegraphics[width=7.0cm, clip] {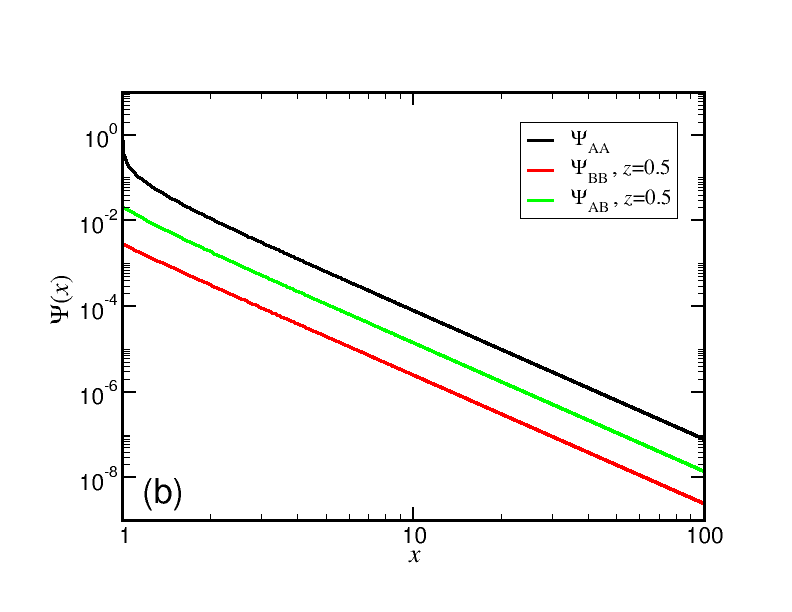}
      \end{minipage}
      \end{center}
\caption{(a) The interaction potentials of
Eq.~(\ref{potentials:eq}) for $z=0.3296$ (full lines) and $z=0.5$
  (broken lines). (b) Double logarithmic plots of the $\Psi_{ij}(x)$ for
  $z=0.5$, demonstrating their $~\sim x^{-3}$ power-law dependence for
  large $x$-values.}
\label{fig_potential}
\end{figure*}

For system I, Sun and Stirner\cite{Sun01} have derived the
following expressions for the pair potentials, $\Phi_{ij}(r)$, acting
between the two species of particles:

\begin{equation}
\Phi_{ij}(r) = \left \{ \begin{array}{ll} \infty & r \leq (R_i + R_j)
                \\ \frac{P_iP_j}{16\pi\epsilon R_i
                R_j}\frac{1}{r}\;\ln \left[\frac{r^2-(R_i-R_j)^2}{r^2
                - (R_i+R_j)^2}\right] & r > (R_i + R_j) \end{array} \right.
~~~~ i,j = \text{A, B}.
\label{potentials:eq}
\end{equation}
Here, $R_i$ and $P_i$ are the radius and the dipolar moment of species
$i$ and $\epsilon$ is the dielectric constant of water. 
In what follows, we assume that $P_i = \alpha R_i^3$, with some
proportionality constant\cite{foot} 
$\alpha$.
Introducing
$z \equiv R_{\rm B}/R_{\rm A} < 1$ and
$x \equiv r/d_{\rm A}$ ($d_{\rm A} = 2\,R_{\rm A}$), and factoring out a prefactor
common to all three interactions $\Phi_{ij}(r)$, we arrive 
at the following expressions
for the dimensionless interaction potentials $\Psi_{ij}(x)$:
\begin{eqnarray}
  \Psi_{\rm {AA}}(x) & = & \frac{1}{x} \ln \left[\frac{x^2}{x^2-1} \right]\qquad\qquad\qquad 
  {\rm for}\,\,x \geq  1; \\
  \Psi_{\rm {BB}}(x) & = & \frac{z^3}{x} \ln \left[\frac{x^2}{x^2-z^2} \right]
  \qquad\qquad\,\,\,\,\,\,\,{\rm for}\,\, x \geq z; \\
  \Psi_{\rm {AB}}(x) & = & \frac{z^{3/2}}{x}
\ln \left[\frac{4x^2-(1-z)^2}{4x^2-(1+z)^2} \right]\,\,\,\,\,\,\,\,{\rm for}\,\, x \geq (1+z)/2.
\end{eqnarray}
The functions $\Psi_{ij}(x)$ are displayed in Fig.\ \ref{fig_potential}(a) for two different values
of $z$. Expanding the logarithms for $x \gg 1$ yields the approximate expressions
$  \Psi_{ij}(x) \cong (z_i z_j)^{5/2}/x^3$, where $z_{i,j} = 1$ or $z$ if $i,j =
{\rm A}$ or ${\rm B}$. 
Fig.\ \ref{fig_potential}(b) shows the asymptotic,
power-law form of the interactions. In two dimensions, where the exponent of the
power-law exceeds the dimension of space, 
the subtle issues of the shape-dependence
of the thermodynamics, which arise for {\it three-dimensional} 
dipoles,\cite{groh:prl:97} are absent.

For system II, on the other hand, where the superparamagnetic colloids
have a susceptibility ratio $m < 1$, the dimensionless interaction
potentials read\cite{maret:prl:06,likos:jpcm:06,assoud:epl:07} as $\bar \Psi_{ij}(x) = m_im_j/x^3$,
where $z_{i,j} = 1$ or $m$ if $i,j = {\rm A}$ or ${\rm B}$. If the overall density of
system I is sufficiently low, so that the asymptotic forms of 
$  \Psi_{ij}(x)$ above hold, systems I and II become equivalent,
with the correspondence $z^{5/2} \leftrightarrow m$. The ground states of system II
have been recently analyzed with conventional methods in the work
of Assoud {\it et al.}\cite{assoud:epl:07} for $0.08 \leq m \leq 1$, being
inspired by the phase diagram of a two dimensional, binary hard-sphere
mixture.\cite{Lik92} Here, we concentrate on system I in the domain $0.1 \leq
z \leq 0.5$, corresponding to much stronger asymmetries (i.e., $3\times
  10^{-3} < m < 0.18$) but still having sufficient overlap in the parameter
range with that of Ref.\ \onlinecite{assoud:epl:07} to allow for quantitative  
comparisons. Finally, the system is characterized by the number-density $\eta$
or its dimensionless counterpart, $\eta d_{\text{A}}^2$.

It is well-known that binary mixtures tend to show a board spectrum of rather
complex alloy phases, most of them being
extremely hard to guess. With the ambition to cover as large a variety of
structures as possible and in an effort to cope with this problem in an efficient way, we
employed genetic algorithms\cite{forrest:science:93} (GAs) instead of following the
conventional approach to the problem of finding minimum energy configurations (MECs),
which relies on preselected sets of candidate structures. 
GAs were invented in the 1970s by Holland {\it et al.} to
solve high dimensional and complex problems in engineering science.\cite{Hol75} They are
optimization techniques, modeled after the natural process of evolution, and
mimic certain biological
mechanisms, such as mating and mutation, to find the optimal solution to a proposed
problem. Due to its special design, a GA is able to take the whole search
space into account at once and at the same time to concentrate its computing
efforts on promising regions. It is this \emph{global scope} that makes GAs an
efficient and widespread tool in fields like economics and engineering. In
the realm of structure optimization, a typical problem in atomic
condensed matter physics, they have found applications in
determining optimal atomic clusters,\cite{ho:prl:95, ho:nature:98} and, very recently,  
in optimizing extended spatial structures.\cite{Oga06,Sie07} In soft matter,
applications of GAs to 
predict equilibrium crystal structures for one-component systems has
already delivered remarkable results, both in two\cite{For07} and
three dimensions,\cite{Got04,Got05Lik,Mla06PRL} but binary systems have
not been looked upon with GAs so far.

Based on the approaches presented in
Ref.\ \onlinecite{Got05Kah}, we designed a method to determine 
ordered MECs for binary monolayers. In our method, lattice parameters
are freely optimized with respect to the free energy, which, at $T=0$, reduces
to the lattice sum $U$ of the ordered structure. The efficiency of the GA allows us to
perform our search for equilibrium structures among all possible lattices,
without posing any bias on the algorithm whatsoever. The only limiting factor
constraining our search is the maximum of particles per cell that the
algorithm can handle in reasonable time. The risk of overlooking relevant
structures at any point in the process is thus minimized.

In general, we have to deal with non-Bravais lattices, 
with $s (>1)$ particles per unit cell. We parametrized these lattices with
$n_{\rm A}$ particles in the unit cell belonging to species A and the
remaining $n_{\rm B}=s-n_{\rm A}$ particles to the other species, B, coding
the position vectors of all particles in a binary fashion.
We obtained results for particle size ratios in the range of $z=0.1$ to
$z=0.5$, including the value of $z=0.3296$ that corresponds to the particle size
ratio used in previous Monte Carlo (MC) simulations.\cite{Sti05} 
The concentration of small particles,
$C={n_{\rm B}}/{s}$, was systematically varied for every size ratio. Most of
the calculations were performed with a maximum of eight particles per unit
cell, so $C$ lied in the range of $1/8 \leq C \leq 7/8$.
Additional calculations were performed for $C=7/9$, as exotic phases
were to be expected for this particular ratio.\cite{Lik92} 

Before presenting 
the results obtained by the GA, let us discuss some generic expected features
of the MECs, arising from the functional form of the
potential. In sufficiently dilute one-component systems, a hexagonal lattice is formed since 
$\Psi_{ij}(r) \sim r^{-3}$ at long distances. This ``default structure''
is also found in two limiting cases of the binary mixture: first, for
$z\rightarrow 0$, when the smaller B-particles are of vanishing size
compared to the larger A-particles, and second for $z\rightarrow 1$,
when the two species become indistinguishable.
In the first case we expect the large particles to form
hexagonal lattices, more or less unperturbed by the presence of the small
ones. If $z$ is small, the B-particles can form independent patterns in
 the interstitial regions of the hexagonal lattice formed by the A-particles 
without paying a high penalty in energy. Thus, distinct groups of small
particles should be observed with their size fulfilling the stoichiometric requirements.
In the opposite case, as the similarity in size increases, the B-particles
require more space. Now the concentration and the dependence of the
interspecies interactions on the distance are the key quantities that
determine the ordered equilibrium structure. 

\begin{figure*}
      \begin{center}
      \begin{minipage}[t]{17.2cm}
      \includegraphics[angle=270,width=13.6cm, clip] {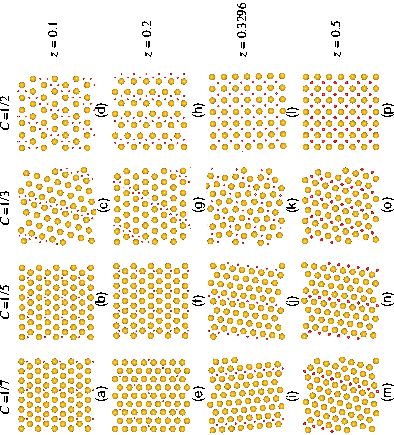}
      \end{minipage}
      \end{center}
\caption{Minimum energy configurations for $C \le 1/2$, {\it i.e.} $n_{\rm B}\leq n_{\rm
 A}$ for $\eta d_{\rm A}^2 = 0.6$. Particle diameters are not drawn to scale.}
\label{fig_moreA_than_B}
\end{figure*}
\begin{figure*}[h]
      \begin{center}
      \begin{minipage}[t]{17.2cm}
      \includegraphics[angle=270,width=17.0cm, clip] {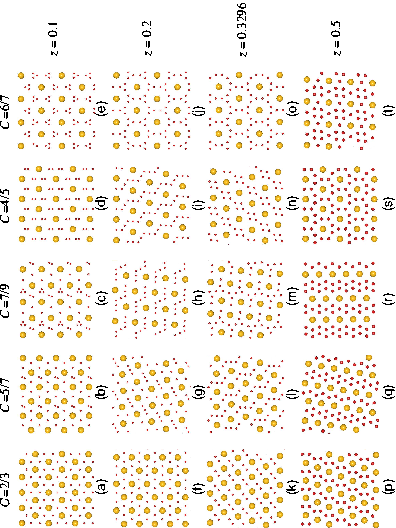}
      \end{minipage}
      \end{center}
\caption{Minimum energy configurations for $C > 1/2$, {\it i.e.} $n_{\rm B}> n_{\rm
  A}$ for $\eta d_{\rm A}^2 = 0.6$. Particle diameters are not drawn to scale.}
\label{fig_moreB_than_A}
\end{figure*}

In Figs.~\ref{fig_moreA_than_B} and \ref{fig_moreB_than_A} we present the
ordered equilibrium configurations found with our GA-method for increasing concentration of small
particles $C$ at an overall particle density of $\eta d_{\rm A}^2=0.6$. 
To facilitate discussion we
divide our results into two blocks: structures with $n_{\rm B} \leq n_{\rm A}$
(Fig.~\ref{fig_moreA_than_B}) and structures $n_{\rm B} > n_{\rm A}$ 
(Fig.~\ref{fig_moreB_than_A}). We begin with the case $n_{\rm B} \leq n_{\rm A}$.
For sufficiently large values of $z$, {\it i.e.}, $z=0.3296$ and $z=0.5$, the small particles are able to perturb
 the (ideal) hexagonal structure formed by the large particles in their immediate surrounding,
even though they represent the minority component. In regions where there are no small particles in the immediate
neighborhood, the hexagonal structure of the large particles prevails, whereas
the small particles are found in the center of squares formed by the large
particles [see Figs.~\ref{fig_moreA_than_B}(i),(j),(m) and (n)]. On the other
hand, if $z$ is below a certain threshold, i.e., $z=0.1$ and $z=0.2$, the
influence of the small particles is not sufficient to cause a substantial
modification of the hexagonal pattern of the large particles. Instead, since
the small particles experience mutually a very weak repulsion, they
tend to stay close together in this $z$-regime, 
arranging in lanes which meander through the hexagonal lattice formed
by the large particles, see Figs.~\ref{fig_moreA_than_B}(c),
\ref{fig_moreA_than_B}(d), \ref{fig_moreA_than_B}(g), and \ref{fig_moreA_than_B}(h).
For very small concentrations, {\it i.e.}, $C=1/7$ and $C=1/5$ [cf. Figs.~\ref{fig_moreA_than_B}(a),
\ref{fig_moreA_than_B}(b), \ref{fig_moreA_than_B}(e), and \ref{fig_moreA_than_B}(f)],  
the `lanes' of the small particles are interrupted by intervening big ones due
to the simple fact that not sufficiently many B-particles are available in the system.
It might be possible that also in these cases pure B-particle lanes will
form but this would require a much larger unit cell, which was not included in
our study.

When the small particles become the majority component,
 the found structures become much
more complex, as can be seen from Fig.~\ref{fig_moreB_than_A}. As 
expected, B-particles are observed to arrange in distinct groups for many
parameter settings. 
We find small particles forming dimers, [Figs.~\ref{fig_moreB_than_A}(b),
\ref{fig_moreB_than_A}(c), and \ref{fig_moreB_than_A}(d)], 
elongated- [Fig.~\ref{fig_moreB_than_A}(g)] and
triangular-trimers [Figs.~\ref{fig_moreB_than_A}(c) and 
\ref{fig_moreB_than_A}(e)], as well as
chain-like pentamers [Fig.~\ref{fig_moreB_than_A}(l)] and
heptamers [Fig.~\ref{fig_moreB_than_A}(h)]. In contrast, the A-particles form
rather simple lattices which accommodate in their interstitial regions these
sometimes rather complex groups of B-particles. For $C=2/3$, dimers of small
particles, observed for moderate values of $z$
[Fig.~\ref{fig_moreB_than_A}(k)] are a precursor of lane formation
[Fig.~\ref{fig_moreB_than_A}(p)]. For small values of $z$, the large
particles form a hexagonal pattern and the small particles are distributed in
the interstitials. At $z=0.3296$ the hexagonal structure of the large
particles is distorted and the small particles are grouped in dimers. If the
particle size ratio is further increased, the dimers of small particles change
their orientation and lane formation sets in which now strongly distorts the
underlying lattice of A-particles. The scenario repeats itself for $C=5/7$,
[vertically from Fig.~\ref{fig_moreB_than_A}(b) to Fig.~\ref{fig_moreB_than_A}(q)]
but here the stoichiometry, which does not accommodate a single B-particle
in the interstitials of A, forces a much richer structure: 
for $z=0.1$, Fig.~\ref{fig_moreB_than_A}(b), the B-particles form an
ordered array of monomers and dimers, which transforms into an array
of monomers and linear trimers for $z = 0.2$, Fig.~\ref{fig_moreB_than_A}(g).
For $z = 0.3296$, the aggregates of B-particles become monodisperse,
zig-zag-like pentamers, Fig.~\ref{fig_moreB_than_A}(l); and, finally, for
$z=0.5$, formation of two interchanging kinds of B-lanes, thick and thin,
takes place, Fig.~\ref{fig_moreB_than_A}(q).

Evidently, the higher the value of $C$, the more surprises are hidden in the system.
Let us discuss the sequences for $C=7/9$, $C=4/5$, and $C=6/7$ in more detail.
For $C=6/7$ and for small $z$-values, the interstitials of the A-lattice
are occupied by {\it three different kinds} of B-aggregates:
monomers, dimers and triangular trimers, Fig.~\ref{fig_moreB_than_A}(c). For
$z=0.2$, B-particles arrange in zigzag-shaped heptamers which distort the hexagonal
pattern of the large particles, Fig.~\ref{fig_moreB_than_A}(h).
However, in contrast to the smaller $C$-values a further increase of $z$ does not directly lead
to lane formation but rather a new, exotic structure intervenes for
$z=0.3296$, Fig.~\ref{fig_moreB_than_A}(m). Here, two
neighboring heptamers merge, forming thereby a sequence of alternately 
oriented cup-like structures, each of them hosting an A-particle. If $z$ is
increased further, thick and thin lanes are again formed,
Fig.~\ref{fig_moreB_than_A}(r). In the second sequence, $C=4/5$, 
the B-dimers [Fig.~\ref{fig_moreB_than_A}(d)], observed for $z=0.1$,
transform into zig-zag lanes [Fig.~\ref{fig_moreB_than_A}(i)], 
pearl-necklace-lanes [Fig.~\ref{fig_moreB_than_A}(n)] and finally into rings, 
each of them surrounding one A-particle, 
[Fig.~\ref{fig_moreB_than_A}(s)], as $z$ grows.
Finally, for the sequence 
$C=6/7$, a ring-like structure is formed for $z=0.1$ to
$z=0.3296$, Figs.~\ref{fig_moreB_than_A}(e), \ref{fig_moreB_than_A}(j) and
\ref{fig_moreB_than_A}(o), where every large particle is surrounded by six
triangular trimers of small particles, forming a structure resembling a Kagome
lattice [Fig.~\ref{fig_moreB_than_A}(e)].
The B-interparticle distance {\it within} the trimers increases with $z$. Finally, for
$z=0.5$ lane formation sets in once more [Fig.~\ref{fig_moreB_than_A}(t)], 
but now the lanes formed by the small particles are interconnected, due to the high concentration of B-particles.

\begin{figure*}[h]
      \begin{center}
      \begin{minipage}[t]{17.2cm}
      \includegraphics[angle=270,width=13.6cm, clip] {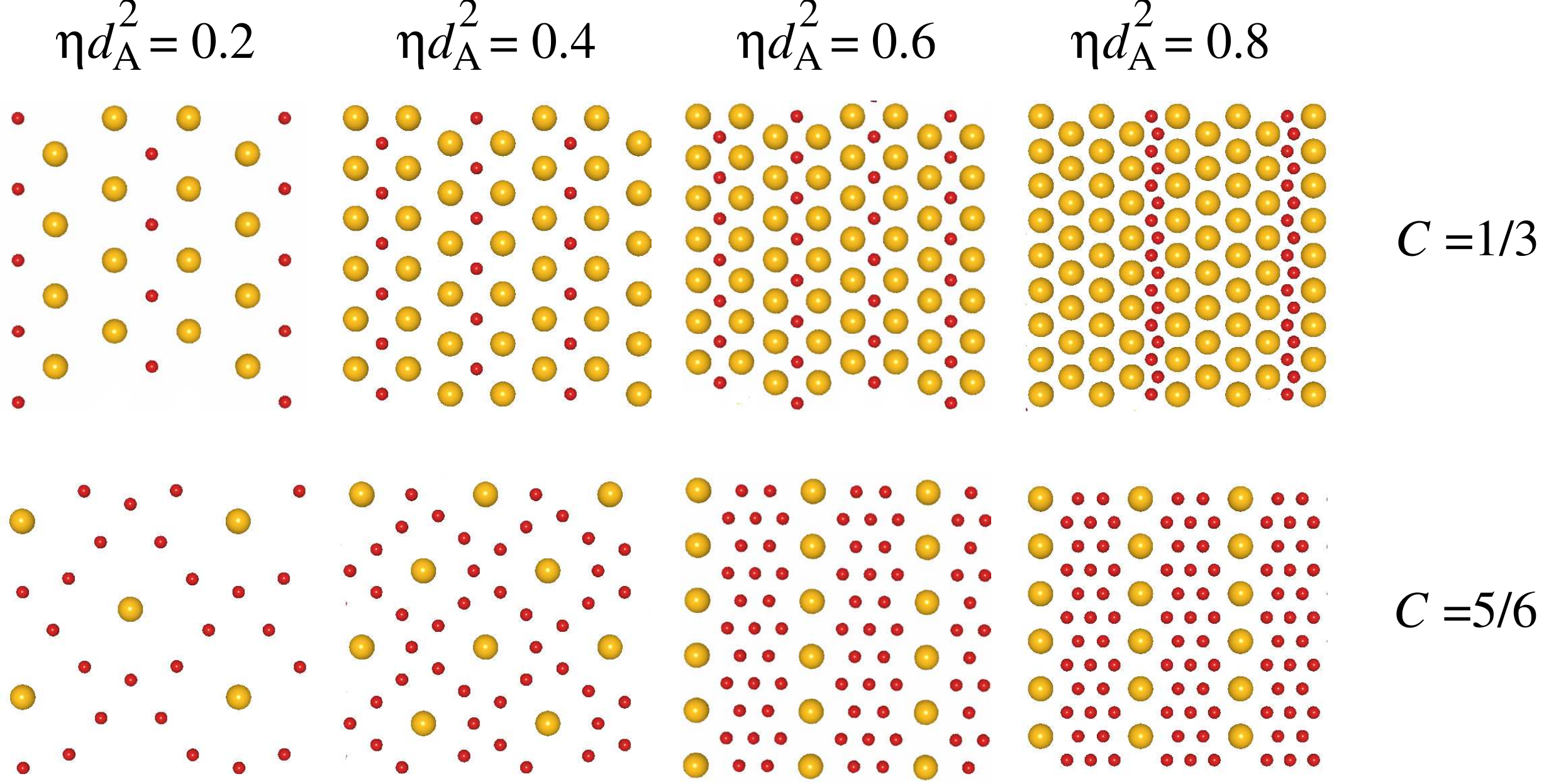}
      \end{minipage}
      \end{center}
\caption{Two sequences of MECs for $z=0.5$ and two different concentrations
  $C$. Structural change is observed in both sequences, once between $\eta
  d_{\rm A}^2=0.6$ and $\eta d_{\rm A}^2=0.8$ (top) 
  and once between $\eta d_{\rm A}^2=0.4$ and $\eta
  d_{\rm A}^2=0.6$ (bottom).}
\label{fig_changing_rho}
\end{figure*}

The structures found by the GA for $z=0.5$ correspond to $m=0.177$ in 
the terminology of Assoud {\it et al.}\cite{assoud:epl:07} For those
stoichiometries that have been considered both in the present work
and in Ref.~\onlinecite{assoud:epl:07},
{\it the same structures} were found by both the conventional approach
and by the GAs. The GAs offer a higher predictive power and flexibility to
identify structures in the regime of large size-disparity and high $C$-values, where the unit cells
become increasingly complex and conventional methods reach their
limitations.\cite{assoud:epl:07,Lik92}

Were the interaction potentials of Eq.~(\ref{potentials:eq}) to be pure power-laws,
as those employed in Ref.~\onlinecite{assoud:epl:07},
the overall density $\eta d_{\rm A}^2$ would be an irrelevant parameter,
in view of the absence of any length scale in the interactions. Since the
logarithmic dependence, as well as the hard cores of the present
interactions, Eq.~(\ref{potentials:eq}),
set in at very small separations [cf. Fig.~\ref{fig_potential}(b)],
we expect that the structures reported in 
Figs.~\ref{fig_moreA_than_B} 
and \ref{fig_moreB_than_A} will be stable for a broad range of small to intermediate
densities. To put this assumption into a check, we have investigated
the ordered equilibrium configurations of our system varying the density
$\eta d_{\rm A}^2$ from 0.2 to 0.8. We report selected results in 
Fig.~\ref{fig_changing_rho}. For concentration $C=1/3$, upper row, the lane-like structure
remains unchanged up to a density at least as high as $\eta d_{\rm A}^2 = 0.6$,
only the distances are scaled as $\eta$ increases. As the density is
further increased to $\eta d_{\rm A}^2 = 0.8$, the following structural
transition takes place: every second B-lane is dissolved by merging one of the
neighboring ones; consequently, the inter-lane distance and the density in
each of these lanes is increased by a factor of two. For $C=5/6$, a structural change induced by density sets in earlier, namely
between $\eta d_{\rm A}^2 = 0.4$ and $\eta d_{\rm A}^2 = 0.6$. Here, the original pattern shows rings of B-particles while, again,
at high densities lane formation is observed.
Moreover, all structures shown here have the lowest energy values found by the
GA but, at the same time, other ones with very small energy differences
have been discovered in some GA runs. Both the effects of the density and the
discussion of `quasi-degenerate' ground states
will be the subject of a future publication.

We have applied genetic algorithms to examine ordered equilibrium
configurations for binary, two-dimensional dipolar mixtures. Despite the simplicity of the
interactions involved, the system shows a tremendous variety and
complexity of the resulting structures. This convincingly 
demonstrates the power of this novel optimization technique to
deal with the problem of finding the global minimum of a rugged
potential energy surface, which becomes increasingly involved
as the number of components of the mixture increases. We have demonstrated that GAs are an effective and reliable new tool which
merits a more widespread appreciation in the soft matter community, on equal
footing with complementary tools, such as Monte Carlo simulated annealing and Molecular Dynamics
simulations. The variety of the encountered structures can easily be
verified experimentally considering either a binary mixture
of two-dimensional superparamagnetic colloidal particles
\cite{maret:prl:06,likos:jpcm:06}
or a binary mixture of polystyrene spheres floating on an oil-water interface\cite{Ave00LAN16a}.

The authors thank Dieter Gottwald for helpful discussions. This work has been supported by the Austrian Research Foundation (FWF), project
numbers P17823-N08 and P19890-N16, and the European Science Foundation
short-visit-grant ``SimBioMa 1730'' (J.F.), by the Marie Curie program of the
European Union, contract number MRTN-CT2003-504712 and 
the Foundation Blanceflor Boncompagni-Ludovisi, n{\'e}e Bildt (F.L.V.),
as well as by the DFG within the SFB-TR6, Project Section C3.
C.N.L.\ wishes to thank the Erwin Schr{\"o}dinger Institute (Vienna),
where parts of this work have been carried out,
for a Senior Research Fellowship and for its hospitality.


\begin{thebibliography}{99}
\bibitem{foot} Another choice is to assume that the dipolar
moment scales with the surface area of the spheres, $P_i = \alpha R_i^2$,
but we focus on $P_i = \alpha R_i^3$ in this work.
\end{thebibliography}
\end{document}